\documentclass[letterpaper,reprint,amsmath,amssymb,aps,nofootinbib,noeprint]{revtex4-2}
\usepackage{graphicx}
\PassOptionsToPackage{normalem}{ulem}
\usepackage{ulem}
\usepackage{color}

\newcommand{\mnras}{Mon. Not. Roy. Astron. Soc.}
\newcommand{\apjl}{Astrophys. J. Lett.}
\newcommand{\jcap}{J. Cosmol. Astropart. Phys.}
\newcommand{\aap}{A\&A}
\newcommand{\aj}{Astron. J.}



\usepackage{babel}
\begin{document}
\title{Probing the Global 21-cm Signal via the Integrated Sachs-Wolfe Effect on the 21-cm Background}
\author{Kyungjin Ahn}
\email{kjahn@chosun.ac.kr}
\author{Minji Oh}
\affiliation{Department of Earth Sciences, Chosun University, Gwangju 61452, Korea}
\begin{abstract}
We propose a novel method to probe the global 21-cm background. This background experiences the integrated Sachs-Wolfe effect (ISW) as the cosmic microwave background does. The 21-cm ISW is modulated by the spectral shape of the global 21-cm signal, and thus the measure of the 21-cm ISW will be a probe of the evolution of the global signal. This strategy naturally mitigates the impact of the Milky Way foreground, which is a common and most significant challenge in conventional 21-cm background probes. With the phase-1 SKA telescope, probing the global 21-cm background would be feasible with a few 1000 hours of observation, enabling consistency checks with existing measures of the global 21-cm signal by EDGES and SARAS that are conflicting with each other.
\end{abstract}

\maketitle
\section{introduction}
\label{sec:intro}
The anisotropy of the cosmic microwave background (CMB) is caused
by various physical processes in the early and the late universe.
Relatively weak yet detected with high significance is the late-time
integrated Sachs-Wolfe effect \cite{Sachs1967} (ISW henceforth) 
on the CMB caused by the temporal variation of the large-scale gravitational
potential in the relatively nearby universe \cite{Fosalba2003,Afshordi2004,Cabre2006,Giannantonio2007,Ilic2011,Giannantonio2012,Collaboration2016,Dong2021,BahrKalus2022}.
In a $\Lambda$CDM universe, after the matter-dominated epoch, the gravitational potential changes
in such a way that photons travel out of the potential well (bump)
with less energy loss (gain) than the energy gain (loss) experienced
while traveling into, causing the net blueshift (redshift) \cite{Sachs1967}.
The measured ISW on CMB, originating from the last scattering surface
 at redshift $z\simeq1000$, thus shows correlation with lower-redshift
($z\lesssim2$) density tracers, or galaxies: the first proposal to
use such a cross-correlation is by \cite{Crittenden1996}. Aside from
the possibility of the curvature density \cite{Kamionkowski1996}
or modified gravity effects \cite{Song2007,Cai2014} for ISW,
the measured ISW by galaxy-CMB correlation is consistent with the present-day
dark energy content in units of the critical density at $\rho_{\Lambda}/\rho_{{\rm crit,0}}\sim0.7$.

Redshifted 21-cm radiation from HI gas at $100\gtrsim z\gtrsim 6$
is another radiation background that originates from high redshift and
will experience the ISW effect. Its sky-averaged brightness temperature, or the global 21-cm background temperature ($\delta T_{b}$($z$) henceforth),
has temporal variation determined by cosmology (baryon
temperature and the collisional excitation of the spin-orbit
angular momentum $J$ of hydrogen) and astrophysics (impact on
$T$ from X-ray heating, the ultraviolet background due to early galaxies
and the corresponding radiational excitation of $J$). Such a temporal evolution is marked by the observed frequency $\nu$ that is related to the originating redshift $z$ through the relation $\nu=1.42\,{\rm GHz}/(1+z)$.

The main channel
for the 21-cm background to be correlated with the low-redshift galaxies
is through the ISW effect (another subdominant channel in \cite{Giannantonio2007}). As will be shown in this paper, the temperature fluctuation
of a frequency-dependent photon background caused by ISW is proportional
to the amplitude of the global value and the spectral feedback:
\begin{equation}
\Delta T(\nu)\propto T(\nu)-\frac{\partial T(\nu)}{\partial\ln\nu}.
\label{eq:DT-T-proportion}
\end{equation}
Therefore, one can in principle measure the cross-correlation between
the 21-cm background in different frequency (redshift) bins and the galaxy distribution at low redshifts, and use the frequency
variation in the cross-correlation to probe $\delta T_{b}(z)$. 

The 21-cm ISW as a measurement of $\delta T_{b}(\nu)-\partial\delta T_{b}/\partial\ln\nu$
can then be used to check consistency with already existing observations
of the global 21-cm signal by monoblock receivers: EDGES (Experiment
to Detect the Global EoR Signature; \cite{Bowman2018}) and SARAS3
(Shaped Antenna measurement of the background RAdio Spectrum, version
3; \cite{SARAS2022}). Both experiments measure the whole sky without
interferometry, and a successful removal of the strong foreground
composed of the Milky Way radio signal, the extragalactic background
and CMB should be performed. These two experiments exhibit mutually conflicting
results: the EDGES signal is reported to have $\delta T_{b}(\text{\ensuremath{\nu\sim80}\,{\rm MHz}})\sim-500\,{\rm mK}$,
while the SARAS3 signal is consistent with a null result and rejects
the EDGES signal at $\sim1.8\sigma$. The absorption trough by the
EDGES is much stronger than what is expected in the $\Lambda$CDM
cosmology and thus has been interpreted as a possible signature for
non-standard cosmology \cite{Barkana2018}. The refutation by SARAS3
seems justified as we have better control of its systematics than of the EDGES instrument \cite{SARAS2022}. However, the ambiguity in the foreground
removal process is quite substantial in these ``monopole-only''
observations \cite{Hills2018} and therefore an independent measurement
of $\delta T_{b}(\nu)$ would be essential for confirmation. Our proposition here is one
of such independent measurements.

Detecting
the 21-cm ISW effect for the purpose of obtaining redundancy to the already-measured CMB ISW effect for cosmology
has been proposed by \cite{Raccanelli2016}. 
We turn around their proposition and use the 21-cm ISW as a probe of $\delta T_{b}(z)$ using the dependence of the ISW on the global temperature (Eq.~\ref{eq:DT-T-proportion}). We define an observable which is based on the 21-cm ISW and designed to probe $\delta T_{b}(\nu)$, different from what was used in \cite{Raccanelli2016}. This will be found practically equivalent to giving up the redundancy in cosmological information but instead gaining the information on the amplitude and the spectral shape of the global 21-cm background.
We will show that such a probe would become practical with the upcoming Square Kilometre Array (SKA) aiming at the 21-cm background anisotropy from the $z\simeq 27$ -- $6$ universe, contrary to the pessimistic forecast by \cite{Raccanelli2016} 
who based the noise estimation at $z=30$ which is too high for the sky noise to be controllable and used an observable different from what we will use in this paper.

The paper is organized as follows. In section~\ref{sec:theoryobs}, we show how the 21-cm background is affected by ISW, and develop an observable based on the 21-cm ISW but aimed at probing $\delta T_{b}(z)$. In section~\ref{sec:result}, we present estimated detectability of the observable with Phase-1 SKA specification to show its encouraging prospect. We also include the detectability when only using superstructures as 21-cm ISW drivers. In section~\ref{sec:discussion}, we discuss the observational strategy to realize the measurement and some caveat in our error estimates. Appendices are dedicated for technical details that were used for estimating observational uncertainties.

\section{theory and observable}
\label{sec:theoryobs}

\subsection{Theory}
\label{sec:theory}
We start from the energy equation of a photon on a line of sight of
an observer, first calculated in \cite{Sachs1967}. All calculations
henceforth are accurate to the linear order. A freely traveling photon
reaching an observer at present experiences the temperature shift
$\Delta T\equiv T-\langle T\rangle $ given by
\begin{align}
\left.\frac{\Delta T(\hat{n})}{\langle T\rangle }\right|_{\eta_{0}}=&\left.\frac{\Delta T(\hat{n})}{\langle T\rangle }\right|_{\eta_{i}}-\Psi(\eta_{0})+\Psi(\eta_{i})+\left.{\bf v}\cdot\hat{n}\right|_{\eta_{i}}^{\eta_{0}}\nonumber\\
&+\int_{\eta_{i}}^{\eta_{0}}2\Psi'(\hat{n},\eta)d\eta,
\label{eq:dTall}
\end{align}
where $T$ is the temperature, $\langle T\rangle $
is the sky-averaged $T$ (monopole), $\eta_{i}$ and $\eta_{0}$ are
the conformal times of some early onset (usually at CMB decoupling)
and the present respectively, $\hat{n}$ is the sky-direction unit
vector, $\Psi$ is the gravitational potential, and $' \equiv\partial/\partial\eta$.
With $\eta$ and $\Psi$, the  Friedmann-Lema\^itre-Robertson-Walker metric perturbed in the linear order reads
\begin{equation}
  ds^{2} = a^{2}(\eta)\left\{ -(1+2\Psi)d\eta^{2} +(1-2\Psi)\delta_{ij}dx^{i}dx^{j} \right\},
\label{eq:FLRW}
\end{equation}
where $a$ is the scale factor, $x^{i}$ is the comoving spatial coordinate, $\delta_{ij}$ is the Kronecker delta,  and only one perturbation term $\Psi$ (as both the curvature perturbation and the gravitational potential) is sufficient when ignoring the anisotropic stress tensor (a valid approximation for the linear ISW: e.g. \cite{Afshordi2004}).
While taking $\eta_{i}$
as the conformal time for the last scattering surface is usual in
the CMB literature, we take $\eta_{i}\equiv\eta(z)$ to generalize
for the background that originates from many different redshifts,
such as the 21-cm background.

We now generalize  Eq.~(\ref{eq:dTall}) but focus only on the late-time ISW
with a frequency-dependent brightness temperature\footnote{One could instead formulate the 21-cm ISW in terms of the flux density or the specific intensity, whose amplitude at frequency $\nu$ (redshift $z$ for spectral-line backgrounds) defines the brightness temperature $T(\nu)$ ($T(z)$) that needs not be a thermal temperature, in general, as in the case of the 21-cm brightness temperature.} field (contrary to the CMB
temperature which is independent of frequency). Also considering the
global optical depth $\tau$ due to Thomson scattering by free electrons
\cite{Giannantonio2012} for a collection of photons, we have
\begin{equation}
\left.\frac{\Delta T(\nu,\hat{n})}{\langle T(\nu)\rangle }\right|_{\eta_{0}}=\int_{\eta_{i}}^{\eta_{0}}e^{-\tau(\eta)}2\Psi'(\hat{n},\eta)d\eta\label{eq:ISWwtauu}
\end{equation}
where $\tau(\eta)\equiv\int_{\eta}^{\eta_{0}}n_{{\rm e}}(\eta)\sigma_{{\rm T}}a(\eta)d\eta$,
with the electron number density $n_{{\rm e}}$, the scale factor $a$  and the Thomson scattering
cross section $\sigma_{{\rm T}}$. In practice, in the $\Lambda$CDM
framework, ISW becomes effective only after $\Omega_{\Lambda}\sim\Omega_{m}$
(or at $z\lesssim2$) and thus the exact value of $\eta_{i}$ does
not matter. The observed frequency, excluding the kinematically
induced dipole \cite{Hotinli2023}, is shifted from the cosmological
one as \footnote{Note the disappearance of $\exp[-\tau(\eta)]$ term compared to Eq.
(\ref{eq:ISWwtauu}). }
\begin{equation}
\nu(\hat{n},z)-\bar{\nu}(z)=\int_{\eta_{i}}^{\eta_{0}}2\Psi'd\eta\label{eq:nu}
\end{equation}
where $\bar{\nu}(z)\equiv\nu_{21}/(1+z)=1420\,{\rm MHz}/(1+z)$, whose $z$-dependence will
henceforth be dropped in notation for simplicity, is the redshifted
21-cm line frequency in an unperturbed universe. 
As shown in \cite{Raccanelli2016}, combining Eqs. (\ref{eq:ISWwtauu})
and (\ref{eq:nu}), 
the observed temperature difference at $\bar{\nu}$ 
becomes
\begin{align}
\Delta T&(\bar{\nu},\hat{n})  \nonumber\\
&=\langle T(\bar{\nu})\rangle\int_{\eta_{i}}^{\eta_{0}}\left(e^{-\tau(\eta)}-\frac{\partial\ln\langle T(\bar{\nu})\rangle }{\partial\ln\bar{\nu}}\right)2\Psi'(\hat{n},\eta)d\eta \nonumber\\
&\simeq \left(\langle T(\bar{\nu})\rangle-\frac{\partial\langle T(\bar{\nu})\rangle }{\partial\ln\bar{\nu}}\right)\int_{\eta_{i}}^{\eta_{0}}2\Psi'(\hat{n},\eta)d\eta,
\label{eq:ISWnubar}
\end{align}
where the spectral feedback term including ${\partial\langle T(\bar{\nu})\rangle}/{\partial\ln\bar{\nu}}$ arises when we Taylor-expand $\Delta T(\nu,\hat{n})$ into terms in $\bar{\nu}$, with the help of Eq. (\ref{eq:nu}). Physically, this feedback occurs because, when an ISW driver (e.g. a potential well) shifts the observed frequency and temperature, the observer probes the temperature fluctuation situated at the other frequency modulated by the same ISW effect.
In comparison, the CMB-only shift becomes (we denote the CMB by ``$\gamma$''
throughout this paper)
\begin{align}
\Delta T_{\gamma,0}(\hat{n})&=\langle T_{\gamma,0}\rangle \int_{\eta_{i}}^{\eta_{0}}e^{-\tau(\eta)}2\Psi'(\hat{n},\eta)d\eta \nonumber\\
&\simeq\langle T_{\gamma,0}\rangle \int_{\eta_{i}}^{\eta_{0}}2\Psi'(\hat{n},\eta)d\eta .
\label{eq:ISWCMB}
\end{align}
In Eqs. (\ref{eq:ISWnubar}) and (\ref{eq:ISWCMB}), we approximated $\tau(\eta)\simeq 0$ because only the late, low-density universe obtains the optical depth in the case of ISW. In the concurrent $\Lambda$CDM model, the ISW-relevant (when $\Psi' \ne 0$) regime limits $\tau(\eta)\lesssim\tau(z\le 2)\simeq 0.0087$, and thus this approximation is justified if one were not to calculate the ISW effect at sub-percent level accuracy. 

The low-frequency global temperature is the one shifted from the global
CMB temperature $\langle T_{{\gamma,0}}\rangle$ ($=2.725\,{\rm K}$) such
that $\langle T(\bar{\nu})\rangle =\langle T_{\gamma,0}\rangle +\delta T_{b}(\bar{\nu})+T_{{\rm FG}}(\bar{\nu})$,
where $\delta T_{b}(\bar{\nu})$ is the global 21-cm brightness temperature
and $T_{{\rm FG}}(\bar{\nu})$ is the foreground temperature by the Milky
Way, extragalaxies, and possibly some unknown foreground (background).
Let us assume for the sake of discussion that $T_{{\rm FG}}$ comes
mostly from the Milky Way and thus does not produce any ISW effect. This will be realized by correlating the temperature field with the galaxy distribution.
Then, one can readily see that the ISW effect on the low-frequency temperature deviates from that of the CMB according to
\begin{equation}
\Delta T(\bar{\nu},\hat{n})=\Delta T_{\gamma,0}(\hat{n})\left( 1+ \frac{\delta T_{21}(\bar{\nu})}{\langle T_{\gamma,0}\rangle}\right) , 
\end{equation}
where 
\begin{equation}
\delta T_{21}(\bar{\nu})\equiv\delta T_{b}(\bar{\nu})-\frac{\partial\delta T_{b}(\bar{\nu})}{\partial\ln\bar{\nu}}.
\label{eq:T21}
\end{equation}
Therefore, measuring the low-frequency ISW effect at varing frequencies
and comparing this to the CMB ISW will enable one to probe $\delta T_{b}(\bar{\nu})$ or more directly $\delta T_{21}(\bar{\nu})$. 

The galaxy-temperature cross-correlation has been the only practical measure of the ISW for CMB, and it will be much more so for the 21-cm due to the weakness of the signal. The cross-correlation angular power spectrum of $\Delta T(\bar{\nu},\hat{n})$ in the radio frequency
and galaxy overdensity $\delta_{{\rm g}}$ at observing frequency
$\bar{\nu}$ is given by
\begin{align}
C_{\ell}^{gT}(\bar{\nu})=&
\left\{ \langle T_{\gamma,0}\rangle+\delta T_{21}(\bar{\nu})\right\} \nonumber\\
&4\pi\int d\ln k\,\Delta_{0}^{2}(k)W_{\ell}^{{\rm g}}(k)W_{\ell}^{{\rm ISW}}(k),
\label{eq:ClgT}
\end{align}
which is a specific case  of the general correlation angular power spectrum of fields $X$ and $Y$ (see Appendix \ref{sec:corr} for details),  
\begin{equation}
C_{\ell}^{XY}(\bar{\nu})=N^{XY}(\bar{\nu})4\pi\int d\ln k\,\Delta_{0}^{2}(k)W_{\ell}^{X}(k)W_{\ell}^{Y}(k),\label{eq:CXY}
\end{equation}
with $X=g$, $Y=T$ and $N^{gT}(\bar{\nu})=\langle T_{\gamma,0}\rangle+\delta T_{21}(\bar{\nu})$. In equations (\ref{eq:ClgT}) and (\ref{eq:CXY}), $W_{\ell}^{X}(k)$ is the window function for a field $X$ relating the wavenumber $k$ of the field to the multipole $\ell$ projected on the sky.  
$\Delta_{0}^{2}(k) \equiv {k^{3}}P_{0}(k)/{2\pi^{2}}$ is the variance of the present-day matter power spectrum $P_{0}(k)$,
\begin{align}
W_{\ell}^{{\rm g}}(k) & =\int dz\,b(z)\phi(z)D(z)j_{\ell}[kr(z)],\nonumber \\
W_{\ell}^{{\rm ISW}}(k) & =\int dz\frac{3\Omega_{m,0}H_{0}^{2}}{k^{2}}\frac{dg}{dz}j_{\ell}[kr(z)],\label{eq:window}
\end{align}
$\phi(z)$ is the normalized galaxy selection function \cite{Giannantonio2012}, $b(z)$ is the linear galaxy
bias, $D(z)$ is the linear growth factor of the density fluctuation,
$j_{\ell}$ is the spherical Bessel function of order $\ell$, $r(z)=\eta_{0}-\eta(z)$
is the comoving distance, $\Omega_{m,0}$ is the matter fraction at
present, $H_{0}$ is the Hubble parameter at present, and $g(z)=D(z)(1+z)$
is the linear growth factor of the gravitational potential \footnote{For a pedagogical derivation of the ISW cross power, see Ref. \cite{Afshordi2004,Afshordi2004a},
which yields Eq. (\ref{eq:window}).}. If the foreground
comes only from the nearby ($z\simeq0$) universe, $C_{\ell}^{gT_{{\rm FG}}}=0$
and thus $C_{\ell}^{gT}(\bar{\nu})\propto\left(\langle T_{\gamma,0}\rangle +\delta T_{21}(\bar{\nu})\right)$ as we put in Eq. (\ref{eq:window}). However, some fraction of the foreground can originate from higher redshift range, which is the subject we discuss in Section~\ref{sec:discussion}.

\subsection{Observable}
\label{sec:observable}
We define an observable as a combination of $C_{\ell}^{gT}(\bar{\nu})$ and $C_{\ell}^{gT_{\gamma}}$. In principle, both quantities will have the identical angular pattern on the sky, because ISWs on the CMB and the 21-cm background are caused by the same gravitational potentials that are probed by galaxies. This enables an observable which converges to a single scalar quantity for any multipole $\ell$, as will be shown in the following. The key forecast will be presented in Fig.~\ref{figure1} and Section~\ref{sec:fs}.

A special, frequency-independent version of Eq. (\ref{eq:ClgT})
with $\delta T_{21}=0$ is the usual high-frequency
CMB one, $C_{\ell}^{gT_{\gamma}}$. It is obvious that we can extract
the global 21-cm signal by defining
\begin{equation}
s_{\ell}(\bar{\nu})\equiv\langle T_{\gamma,0}\rangle \left(\frac{C_{\ell}^{gT}(\bar{\nu})}{C_{\ell}^{gT_{\gamma}}}-1\right)=\delta T_{21}(\bar{\nu}).\label{eq:Clratio}
\end{equation}
Note that $s_{\ell}(\bar{\nu})$ turns out to be independent of $\ell$. The cumulative signal
\begin{equation}
S(\bar{\nu})\equiv\frac{\sum_{\ell=\ell_{{\rm min}}}^{\ell_{{\rm max}}}s_{\ell}(\bar{\nu})}{(\ell_{{\rm max}}-\ell_{{\rm min}}+1)}=\delta T_{21}(\bar{\nu})\label{eq:Clsumratio}
\end{equation}
will maximize the net sensitivity with many harmonics, where we take
$\ell_{{\rm min}}=2$ and $\ell_{{\rm max}}=100$. In practice, $\ell_{{\rm max}}\sim 20$ is already large enough to saturate SNR of $S(\bar{\nu})$, as will be seen in Fig. \ref{fig:SNRdensity}. The cumulative
SNR for $S(\bar{\nu})$ is given by\footnote{This is a rough estimate; a more accuarate error estimate should be
taken for the effect of the masking at least. In practice, $C_{l}^{T_{\gamma}T_{\gamma}}$ measured by the Planck telescope is limited only by cosmic variance at $\ell\lesssim 1100$, and thus we may neglect the detector noise in Eq. (\ref{eq:SNR}).} (deriviation given in Appendix~\ref{sec:fsnoise})
\begin{widetext}
\begin{equation}
\left(\frac{{\rm S}}{{\rm N}}\right)^{2}(\bar{\nu})
=\sum_{\ell=\ell_{{\rm min}}}^{\ell_{{\rm max}}}\left(\frac{\delta T_{21}(\bar{\nu})}{\sigma_{s_{\ell}}(\bar{\nu})}\right)^{2}
=\frac{\delta T_{21}^{2}(\bar{\nu})}{\langle T_{\gamma,0}\rangle^2}
\sum_{\ell=\ell_{{\rm min}}}^{\ell_{{\rm max}}}
\left[\frac{\left(C_{\ell}^{gg}+\bar{n}_{g}^{-1}\right)\left\{ C_{\ell}^{\delta T_{b}\delta T_{b}}(\bar{\nu})+\epsilon_{\ell,T}(\bar{\nu})+\left(\frac{\delta T_{21}(\bar{\nu})}{\langle T_{\gamma,0}\rangle }\right)^{2}C_{\ell}^{T_{\gamma}T_{\gamma}}\right\} }{(C_{\ell}^{gT_{\gamma}})^{2}f_{{\rm sky}}(2\ell+1)}\right]^{-1},
\label{eq:SNR}
\end{equation}
\end{widetext}
where $f_{{\rm sky}}$ is the sky coverage, $\bar{n}_{g}$ is the
surface number density of ISW galaxies (total number of galaxies per steradian), $\epsilon_{\ell,T}$
is the 21-cm systematic error due to the sky brightness given as
\begin{equation}
\epsilon_{\ell,T}=\frac{2\pi\lambda^{2}(\bar{\nu})\left\{ 180\left({\bar{\nu}}/{180\,{\rm MHz}}\right)^{-2.6}\,{\rm K}\right\} ^{2}}{f_{{\rm cov}}^{2}D^{2}\Delta t\Delta\nu}\label{eq:epsilon}
\end{equation}
with the observing wavelength $\lambda$, the maximum baseline $D$ of the densely spaced core of interferometers,
the observing time $\Delta t$, the land coverage $f_{{\rm cov}}$,
and the bandwidth $\Delta\nu$ \cite{Zaldarriaga2004}. We have nomalization coefficients (Eq. \ref{eq:CXY}) $N_{\ell}^{\delta T_{b}\delta T_{b}}=\delta T_{b}^{2}(\bar{\nu})$,
$N_{\ell}^{gT_{\gamma}}= T_{\gamma,0}$ and $N_{\ell}^{T_{\gamma}T_{\gamma}}= T_{\gamma,0}^{2}$.
Especially, $C_{\ell}^{\delta T_{b}\delta T_{b}}$ is dominated by the primordial power spectrum with negligible contribution from the auto-correlation of the 21-cm ISW (Appendix \ref{sec:fsnoise}), and thus the net power $C_{\ell}^{\delta T_{b}\delta T_{b}}$ does not get boosted by the spectral feedback term $\partial\delta T_{b}(\bar{\nu})/\partial\ln\bar{\nu}$. This is an important fact that renders the noise controllable during the epoch when the noise is dominated not by $\epsilon_{\ell,T}$ but by $C_{\ell}^{\delta T_{b}\delta T_{b}}(\bar{\nu})$. The noise on $s_{\ell}(\bar{\nu})$, or $\sigma_{s_{\ell}(\bar{\nu})}$,
depends on the spectral shape of $\delta T_{b}(\bar{\nu})$ and thus
SNR depends on cosmology and astrophysics. Note that one should also consider the noise $\sigma_{\langle T_{\gamma,0}\rangle}$ in the global CMB temperature due to our definition of the signal (Eq. \ref{eq:Clratio}); however, we find it negligible already with the FIRAS observation giving $\sigma_{\langle T_{\gamma,0}\rangle}=0.57\,$mK \cite{Fixsen:1996nj,Fixsen2009} while the net noise $S(\bar{\nu})/({\rm S/N})(\bar{\nu})$ hardly reaches such a small value, and always residing at $> 20\,$mK with all the observational configurations considered in this paper.

\section{Result}
\label{sec:result}
\subsection{full-sky observation}
\label{sec:fs}
We test three distinctive 21-cm models and estimate (S/N)($\bar{\nu}$) given in Eq. (\ref{eq:SNR}).
We
consider two different strategies in measuring $C_{\ell}^{gT}$:
(1) the usual full-sky ISW measurement and (2) the ISW measurement
only through superstructures \cite{Granett2008,Nadathur2012,Flender2013,Collaboration2016,Nadathur2016,Kovacs2019,Hang2021,Kovacs2022a}, to be described in Section~\ref{sec:ss}.
The test models are (1) ``V'' (vanilla), a typical  $\Lambda$CDM
model with galaxies without much feedback on their formation \cite{Furlanetto2006}, (2)
``SRII'' (self-regulated type II), a model under the $\Lambda$CDM
framework but with strong self-regulation of galaxy formation during
the cosmic dawn \cite{Ahn2012,Ahn2021}, and (3) ``NS'' (non-standard):
a model beyond $\Lambda$CDM that has a large dip of $\delta T_{b}\simeq-500\,$mK
around $\nu\simeq80\,$MHz, having resemblance to the EDGES signal.
All these models have $\delta T_{b}>0$ at $z\lesssim9$ with steep
upturns at higher frequencies, which is believed to be natural with moderate X-ray heating
and consistent with what is indicated by the upper limits on the 21-cm
fluctuation put by some of the SKA precursors \cite{Trott2020,Ghara2020a,HERA2023}.
Therefore, all these models have epochs with steep slopes in $\delta T_{b}(\bar{\nu})$
such that $\left|\partial\delta T_{b}/\partial\ln\bar{\nu}\right|\simeq[2-8]\left|\delta T_{b}\right|$
in some range of frequencies and the net signal $\delta T_{21}$ can
be greatly amplified from $\delta T_{b}$. 

\begin{figure*}[t]%

\includegraphics[width=0.65\textwidth]{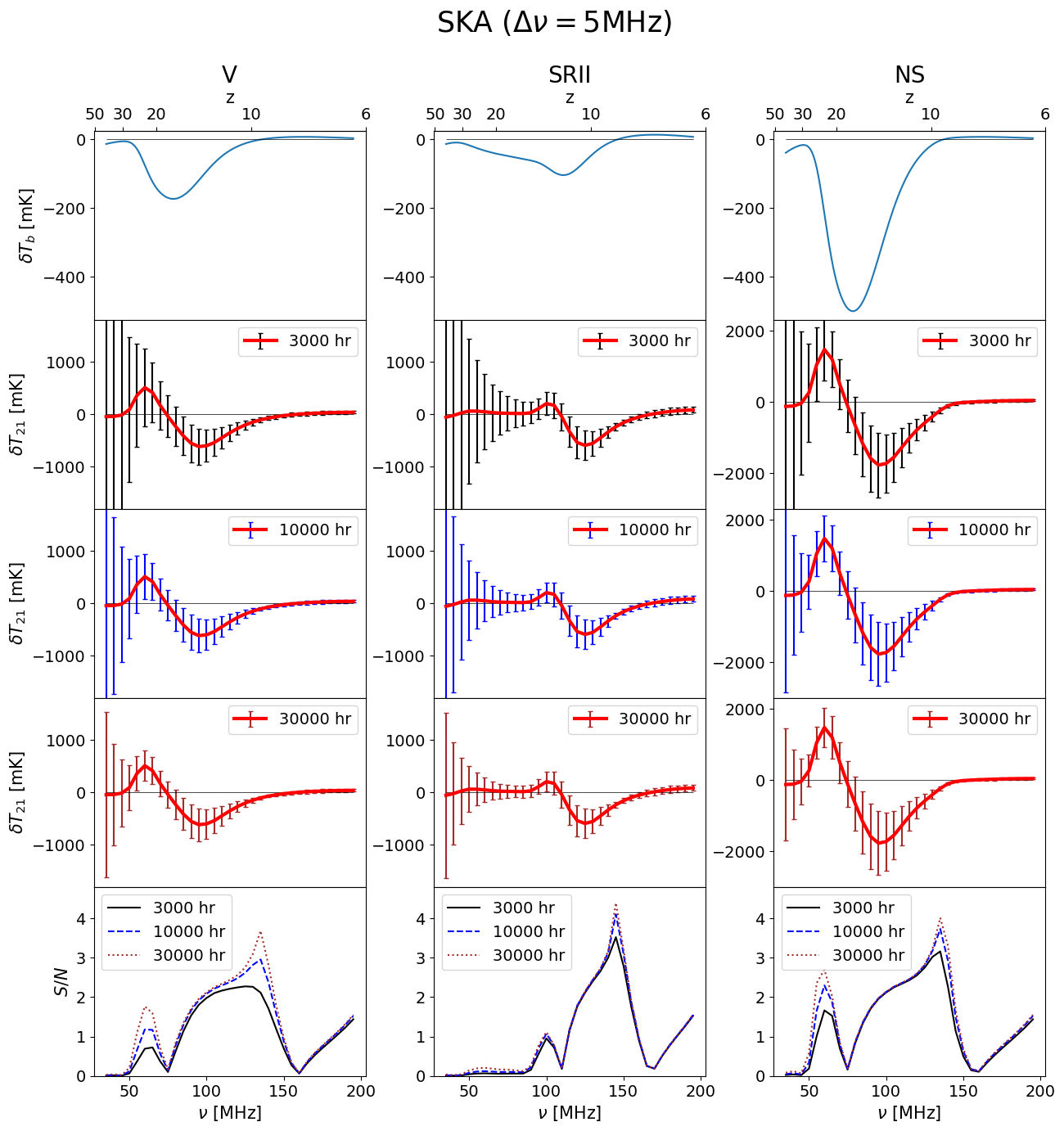}

\caption{Forecasts for measuring the global 21-cm background $\delta T_{21}(\nu)=\delta T_{b}(\nu)-\partial\delta T_{b}/\partial \ln \nu$, through observing the 21-cm ISW effect by the SKA1-LOW telescope with 3000-, 10000- and 30000-hour observing times and 5-MHz bandwidth on 70\% of the sky. Mock signals from three different test models (V: left, SRII: middle, NS: right) are used. From top to bottom, subpanels show the spectra of $\delta T_{b}$ (top), spectra of the observable (thick, red), $\delta T_{21}$, with 1-$\sigma$ error bars when $\Delta t$=3000 (2nd from top), 10000 (middle) and 30000 (4th from top) hours, and finally SNRs (bottom). Error bars and SNR curves follow the same line-color convention (3000 hr: black; 10000 hr: blue; 30000 hr: blown). While there is model dependency, moderate to high-significance detection of $\delta T_{21}$, and thus $\delta T_{b}$, from the cosmic dawn and the epoch of reionization is feasible, with the exception for the shallow-slope (in $\delta T_{b}$) case of SRII at $z\gtrsim 13$ ($\nu \lesssim 100\,$MHz). Observation during the dark ages ($z\gtrsim 30$) remains impossible with our strategy in all cases, due to the ever increasing sky brightness as frequency decreases. 
\label{figure1}}
\end{figure*}%

We take the following observational
parameters that are suitable with the $\sim$1-km core region of the
Square Kilometre Array: $D=1$km, $f_{{\rm cov}}=0.25$, $\Delta t$=\{3000, 10000, 30000\} hr,
$f_{{\rm sky}}=0.7$, $\Delta\nu=5\,$MHz, and galaxy surveys satisfying: $\bar{n}_g\ge (b/2)^{-2}\times 10^{7}{\rm sr}^{-1}$ where $b\equiv \langle b(z)\rangle$ is the average galaxy bias. The last condition and the large $f_{\rm sky}$ requirement are already met with the Wide-Field Infrared Survey Explorer (WISE) galaxy catalogues \cite{Yan2013,Collaboration2016}, and will get more surpassed by the upcoming full-sky galaxy survey Spectro-Photometer for the History of the Universe, Epoch of Reionization, and Ices Explorer (SPHEREX) \cite{Dore2014}. We find that uncertainty due to the galaxy shot noise starts to take effect, mostly at $\bar{\nu}\gtrsim 160$MHz, when $\bar{n}_g$ fails to satisfy the above criterion.

The prospect of using the observable $S(\bar{\nu})$ as a probe of
the global 21-cm signal is encouraging (Fig. \ref{figure1}). Due to the non-exixtence of the correlation between the Milky Way (dominating the radio sky) and the 21-cm background, our method is free from the ambiguity in the foreground removal \footnote{The impact by extragalactic and unknown foregrounds still exist, which is addressed in Section~\ref{sec:discussion}.}. Except for the
dark ages ($z\gtrsim30$; $\bar{\nu}/{\rm MHz}\lesssim46$), we can
achieve reasonably high SNR in probing both the cosmic dawn ($15\lesssim z\lesssim30$;
$89\gtrsim\bar{\nu}/{\rm MHz}\gtrsim46$) and the epoch of reionization
($6\lesssim z\lesssim15$; $202\gtrsim\bar{\nu}/{\rm MHz}\gtrsim89$)
only with the exception of SRII having low SNR in the cosmic dawn.
Especially, the NS case has $\gtrsim(2-3)\sigma$
detectability per frequency band, and thus the 21-cm ISW observation would either rule out or detect the non-standard universe with relatively high significance (compared to the expected theoretical maximum SNR for CMB ISW \cite{Crittenden1996,Afshordi2004}, SNR$\sim7-9$, this can be considered high significance).  
The 21-cm ISW observation would thus place a significant constraint on the global 21-cm background, and
allow consistency checks with the previous, mutually conflicting observations of $\delta T_{b}(\bar{\nu})$  by the EDGES and the SARAS.

\begin{figure}
\includegraphics[width=0.4\textwidth]{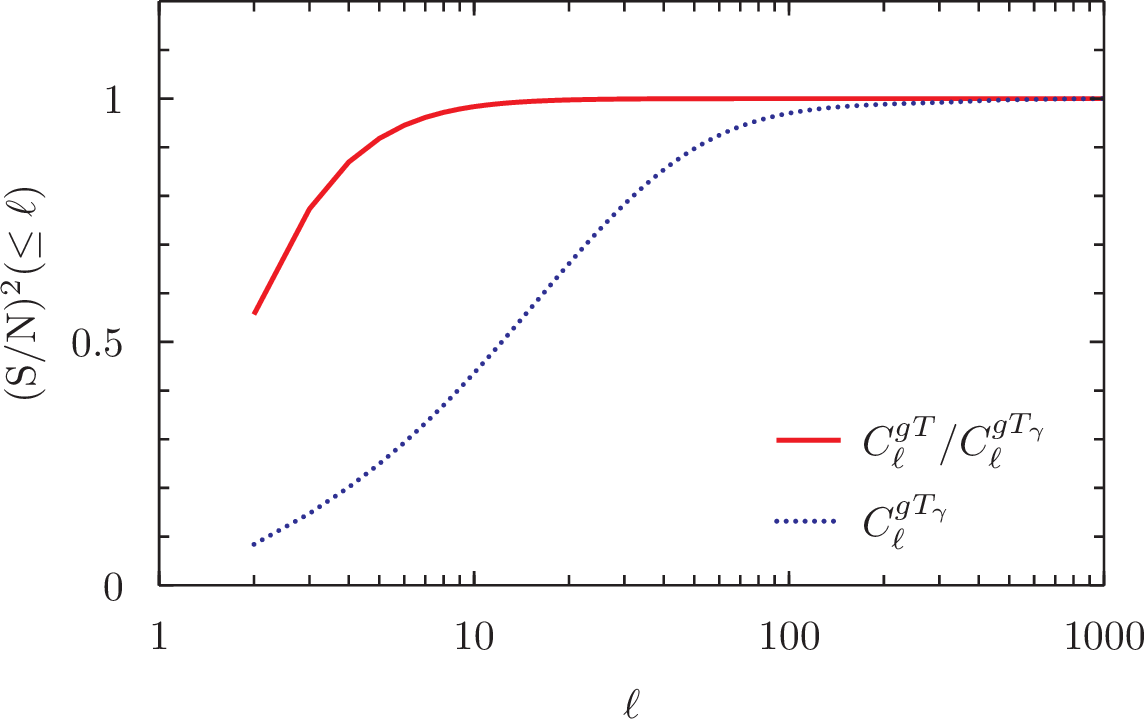}
\caption{Cumulative (S/N)$^2(\le \ell)\equiv \sum_{\ell'=2}^{\ell}({\rm S/N})^2(\ell')$ for the 21-cm probe $S(\bar{\nu})$ at a representative frequency $\bar{\nu}=90\,{\rm MHz}$ (red, solid) and for the CMB-ISW cross-correlation $C_{\ell}^{gT_\gamma}$ (blue, dotted). Curves are normalized to show relative importance of modes for the two cases, such that the saturated $({\rm S/N})^2$ reaches 1 in both cases. Most of the contribution comes from modes at $\ell\lesssim 20$ in the 21-cm probe, while from modes at $\ell\lesssim 200$ in the CMB-ISW cross-correlation. Therefore, it is necessary to accurately probe the largest scales of the sky for the 21-cm-ISW correlation, and also for the CMB-ISW correlation because the net observable is the combination of the two cross-correlations.
\label{fig:SNRdensity}}
\end{figure}

\subsection{superstructure observation}
\label{sec:ss}
Can we use the superstructures (superclusters and supervoids) as the 21-cm ISW driver and probe the global 21-cm background? We find that such a strategy is impractical even with planned telescopes at the moment. The superstructure ISW effect on CMB has been found to be about $A_{\rm ISW}\simeq 1.5-5$, where $A_{\rm ISW}$ is the ratio of the observed $\Delta T$ in superstructures to the $\Delta T$ expected in $\Lambda$CDM due to ISW.   Recent analyses such as \cite{Nadathur2016,Hang2021} with carefully chosen convolution filters, which are used to identify and characterize superstructures, find that ISW in these regions is only marginally amplified, or $A_{\rm ISW}\sim 1.5$, and thus consistent with $\Lambda$CDM. Analyses beforehand using a rather special convolution filter used to find $A_{\rm ISW}\sim 5$, too large to be due to the ISW in $\Lambda$CDM \cite{Granett2008,Nadathur2012,Flender2013,Collaboration2016,Kovacs2019,Kovacs2022a}. Whether or not the superstructure ISW is consistent with the standard cosmology, we may use these as the 21-cm ISW drivers. 

\begin{figure}[t]%
\includegraphics[width=0.45\textwidth]{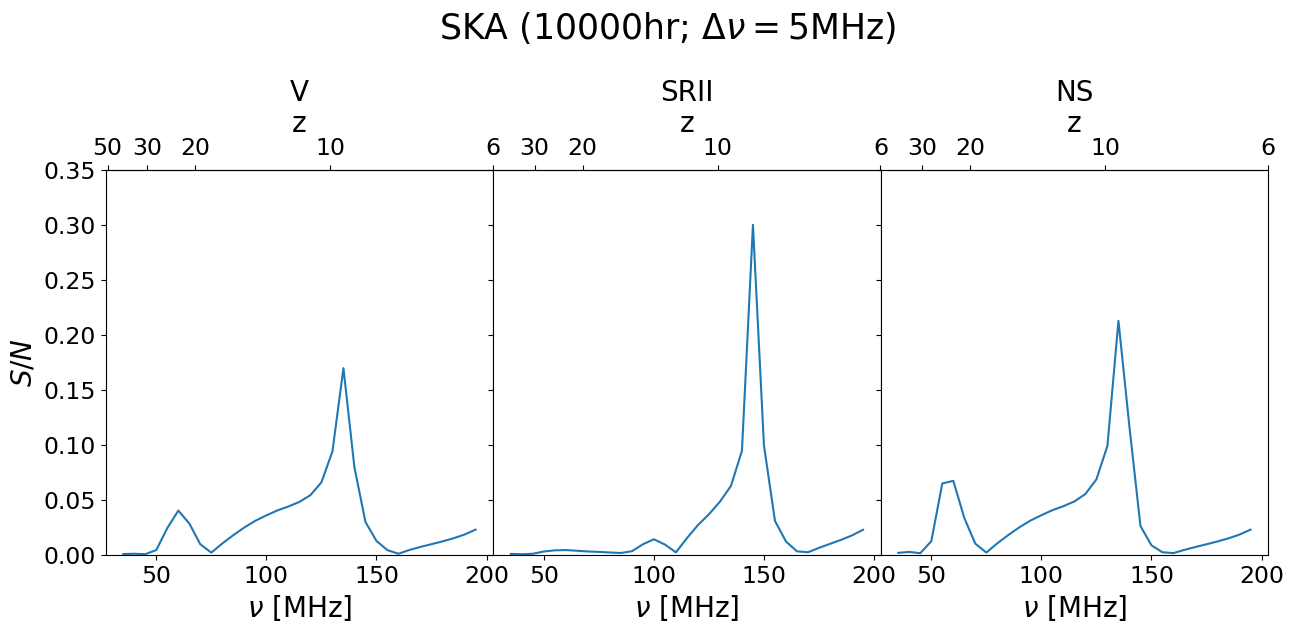}
\caption{Forecasts for measuring the global 21-cm background $\delta T_{21}(\nu)$ via stacked 21-cm ISW effect on $\sim 200$ superclusters, with $\Delta nu=5\,{\rm MHz}$ and observing time of 10000 hours by SKA. In all the tested cases with SKA, SNRs are too low for such observations to be practical, leaving the full-sky observation (Fig. \ref{figure1}) the only viable strategy.
\label{fig:ssSNR}}
\end{figure}%

The 21-cm ISW by superstructures will have
\begin{equation}
\Delta T^{\rm ISW} (\bar{\nu})= A_{\rm ISW} \frac{\delta T_{21}(\bar{\nu})}{T_{\gamma,0}} \Delta \tilde{T}^{\rm ISW}_{\gamma} ,
\label{eq:super21}
\end{equation}
with the theoretically expected $\Delta {\tilde{T}}^{\rm ISW}_{\gamma}\simeq 2\,\mu {\rm K}$ on superstructures in $\Lambda$CDM \cite{Granett2008,Nadathur2016} and comparing this signal to the expected noise inside superstructures, we obtain SNR which is well below 1 in the full frequency range (Fig. \ref{fig:ssSNR}; calculation of SNR described in Appendix~\ref{sec:ssnoise}). This is so even when we use $A_{\rm ISW}=5$. Note that the signal is the average of the stacked $\Delta T^{\rm ISW} (\bar{\nu})$, where we assumed stacking of 200 superstructures that is comparable to the number of superstructures used in \cite{Granett2008,Nadathur2012} to reduce the noise in measuring the CMB ISW.

\section{Discussion}
\label{sec:discussion}
We briefly remark on the impact of the extragalactic foreground and some possible unknown foreground. As stressed earlier, we have null contribution from any radiation of the Milky Way origin, or $C_{\ell}^{gT^{\rm MW}}(\bar{\nu})=0$. Even so, foregrounds of other origin can be problematic. First, extragalactic foreground can be correlated with the late-time galaxy distribution simply because radio emission from these galaxies exist. Therefore, 
\begin{equation}
C_{\ell}^{gT}(\bar{\nu})=C_{\ell}^{gT^{\rm exgal}}(\bar{\nu})+C_{\ell}^{gT^{\rm unknown}}(\bar{\nu})+C_{\ell}^{g\delta T_b}(\bar{\nu})
\end{equation}
in general. Further, both these foregrounds can be decomposed into the intrinsic and ISW-driven parts, because if a part of radiation comes from high redshift, $z\gtrsim 2$, that part of the foreground can also experience the ISW by galaxies at $z\lesssim 2$. The same logic will apply to some unknown foreground. Therefore, in practice, removal of extragalactic and unknown foregrounds from the field of $C_{\ell}^{gT}(\bar{\nu})$ will still be necessary. We will further investigate this issue in a future study.

According to what is shown in Section~\ref{sec:result}, we conclude that the full-sky 21-cm ISW measurement by radio interferometers
is the only practical way to probe the global 21-cm background through ISW. Observing the kinematically induced multipole moments of the 21-cm background anisotropy due to the motion of the earth is another independent measure of the global 21-cm background as we suggested elsewhere \cite{Deshpande2018,Hotinli2023}. There are pros and cons of these two observational strategies, which are complementary to each other. The 21-cm ISW observation does not suffer from the imperfect removal of the Milky Way foreground, thanks to the null cross-correlation with the 21-cm background, while requires interferometric telescopes still in the future with very high sensitivity, such as SKA and HERA. The multipole (practically dipole and possibly quadrupole at best) observation requires much smaller telescopes, while the foreground removal process in each multipole is still required.

For observation of the 21-cm ISW, both HERA (with a strong limitation described below though) and SKA1-LOW (phase-1 SKA in low frequency) meet the thermal noise requirement. However, SKA1-LOW will be the only choice at the moment. HERA is a telescope with $D=300\,{\rm m}$ and $f_{\rm cov}\simeq 0.75$ for the collecting area in the core region to reach $51000\,{\rm m}^2$. This is roughly the same as what we assumed in this paper for SKA in terms of $Df_{\rm cov}$, but the 21-cm ISW observation with HERA would become practical only if the full-sky observation can be performed with the addition of the beam-forming capability. This is because the largest-scale modes should be probed and also the sky coverage is essential to achieve high SNR (Eq. \ref{eq:SNR}). Currently, HERA is planned to observe $1440$-deg$^2$ sky which is far too small for the ISW observation. The core region of SKA1-LOW is what we have assumed as the SKA-core configuration in this paper, planned to be spread over the circular area of $D\simeq 1\,{\rm km}$ with the total collecting area $\sim 215000\,{\rm m}^2$, or $f_{\rm cov}\simeq 0.27$. We even find that 3000-hr integration would still give moderate SNR at $\nu\gtrsim 90\,$MHz. Therefore, the predicted SNR of this paper will become possible with $\gtrsim$1-year operation of SKA1-LOW, which practically amounts to about $\gtrsim 2000$--$4000$-hr observation time depending on the observing condition throughout the total observing time $\Delta t$. With the expected completion of SKA1-LOW construction in year 2029, probing the global 21-cm background through the 21-cm ISW observation could be realized within about a decade from now.

We note that the currently planned observational strategy for SKA1-LOW is likely to be unsuitable for the full-sky 21-cm ISW observation we propose here. The survey area on the sky for SKA1-LOW 21-cm background observation is expected to be limited, to gain very high sensitivity for scales of common interest in the EoR science: $k\sim 0.1$--$1$~Mpc$^{-1}$ \cite{KoopmansSKAwhites2015}. However, our proposal is quite promising in terms of both sensitivity and mitigated foreground effect (that of the Milky Way especially). Therefore, we suggest that our proposal for the low-angular-resolution, full-sky ($\ell \lesssim 20$) observation be given serious consideration as a new science potential of the SKA.

\acknowledgments
KA is supported by NRF-2021R1A2C1095136 and a research grant from Chosun University (2018).

\appendix
\section{angular power spectrum of auto- and cross-correlation}
\label{sec:corr}

For theoretical estimates of the signals and noises, the angular power
spectra of the auto- and cross-correlation are required. The angular
power spectrum of the cross-correlation between fields $X$ and $Y$
is given by 
\begin{equation}
C_{\ell}^{XY}(\bar{\nu})=N^{XY}(\bar{\nu})4\pi\int d\ln k\,\Delta_{0}^{2}(k)W_{\ell}^{X}(k)W_{\ell}^{Y}(k),\label{eq:CXYapp}
\end{equation}
where $N^{XY}$ is the coefficient that renders $C_{\ell}^{XY}$ to
be dimensional when a temperature field is involved, and the frequency
dependence in $C_{\ell}^{XY}$ and $N^{XY}$ appears when the 21-cm
background is involved. Window functions $W_{\ell}(k)$ are shown in
Eq. (\ref{eq:window}) of the main text. This dimensional definition of $C_{\ell}^{XY}$
is convenient as we use two different temperature fields (CMB and
21-cm background), and the 21-cm background experiences the spectral
feedback by the ISW while the CMB does not. Eq. (\ref{eq:CXYapp})
can be applied for the auto-correlation of the galaxy overdensity,
and of the ISW-affected temperature fields. $N^{XY}$'s are given
by 
\begin{equation}
N^{gg}=1;\,\,N^{gT_{\gamma}}=T_{\gamma,0};\,\,N^{g\delta T_{b}}(\bar{\nu})=\delta T_{21}(\bar{\nu}).\label{eq:NXY}
\end{equation}

In both backgrounds, the temperature autocorrelations $C_{\ell}^{T_{\gamma}T_{\gamma}}$
and $C_{\ell}^{\delta T_{b}\delta T_{b}}$ are dominated by the primordial
fluctuations, not by the ISW-driven fluctuations \citep{Giannantonio2012}.
We obtain $C_{\ell}^{T_{\gamma}T_{\gamma}}$ using the Boltzmann solver
``Code for Anisotropies in the Microwave Background'' (CAMB; \citep{Lewis2011}).
With a reasonable assumption of linearity, we can still use Eq.
(\ref{eq:CXYapp}) to obtain $C_{\ell}^{\delta T_{b}\delta T_{b}}(\bar{\nu})$
using the matter density fluctuation: 
\begin{align}
N_{\ell}^{\delta T_{b}\delta T_{b}}(\bar{\nu}) & =\delta^{2}T_{b}(\bar{\nu}) ,\nonumber \\
W_{\ell}^{\delta T_{b}}(\bar{\nu}) & =\int^{\Delta z(\bar{\nu})}dz\,D(z)j_{\ell}[kr(z)]/\Delta z(\bar{\nu}),\label{eq:CdTbdTb}
\end{align}
where $\Delta z(\bar{\nu})$ is the redshift range defined by the
given frequency bandwidth around $\bar{\nu}$, and in effect $W_{\ell}^{g}$ (Eq.~\ref{eq:window})
is used with $\phi(z)=1/\Delta z$ and $b(z)=1$. Because at high redshift,
very small wavenumbers $k\sim(\ell+1/2)/r(z=30\text{--}6)\lesssim0.002\text{--}0.003\,h\,{\rm Mpc}^{-1}$
become relevant in the range of harmonics relevant to the 21-cm ISW, $\ell\lesssim 20$ (see Fig. \ref{fig:SNRdensity}),
and thus the nonlinear effects from the photon-intensity fluctuations
and the inhomogeneous H II structure do not count much in calculating
$C_{\ell}^{\delta T_{b}\delta T_{b}}(\bar{\nu})$ \citep{2005ApJ...624L..65B}.
As stressed in the main text, it is important to notice that $N_{\ell}^{\delta T_{b}\delta T_{b}}(\bar{\nu})$
is equal not to $\delta^{2}T_{21}(\bar{\nu})$ but to $\delta^{2}T_{b}(\bar{\nu})$,
because otherwise the noise would have been uncontrollably large.

\begin{figure}
\includegraphics[width=0.4\textwidth]{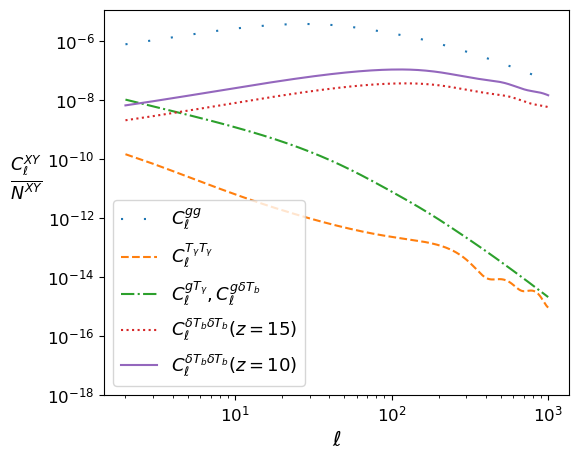}

\caption{Dimensionless auto- and cross-correlation power spectrum of the galaxy
(``g''), the CMB temperature (``$T_{\gamma}$'') and the 21-cm
temperature (``$\delta T_{b}$'').\label{fig:cross-corr}}

\end{figure}

We use $b(z)=2$ and a smooth selection function \citep{Giannantonio2012}
\begin{equation}
\phi(z)=\frac{1}{\Gamma\left(\frac{\alpha+1}{\beta}\right)}\beta\frac{z^{\alpha}}{z_{0}^{\alpha+1}}\exp\left[-\left(\frac{z}{z_{0}}\right)^{\beta}\right]\label{eq:phi}
\end{equation}
with $\alpha=3.457$, $\beta=1.1$, $z_{0}=0.3$, peaking at $z\simeq0.8$.
We experimented with different values of \{$\alpha,$ $\beta$, $z_{0}$\}
and find that our error estimation is hardly affected by these quantities.
The dependence on $b(z)$ roughly appears as the limit we find, $\bar{n}_{g}\ge(b/2)^{-2}10^{7}\,{\rm sr}^{-1}$,
that saturates the SNR at the maximum values we find for values of
\{$\alpha,$ $\beta$, $z_{0}$\} above (Fig. 1 in the main text). Note that the observable $S(\bar{\nu})$ is hardly affected by $b(z)$, because its dependence cancels out when the division ${C_{\ell}^{gT}(\bar{\nu})}/{C_{\ell}^{gT_{\gamma}}}$ is made.
This limit on $\bar{n}_{g}$ might be relaxed if $\phi(z)$ were peaked
at lower redshifts than we have chosen here, due to the relative increase
in $C_{\ell}^{gg}$. We have calculated the integration in $W_{\ell}^{X}$'s
by using the ``FFTLog'' algorithm developed elsewhere \citep{Fang2020},
instead of using the Limber approximation. Fig. \ref{fig:cross-corr} shows dimensionless
auto- and cross-correlations $C_{\ell}^{XY}/N^{XY}$ we obtained.

\section{Noise in observables}
\label{sec:noise}
\subsection{full-sky 21-cm ISW}
\label{sec:fsnoise}

\begin{figure}
\includegraphics[width=0.45\textwidth]{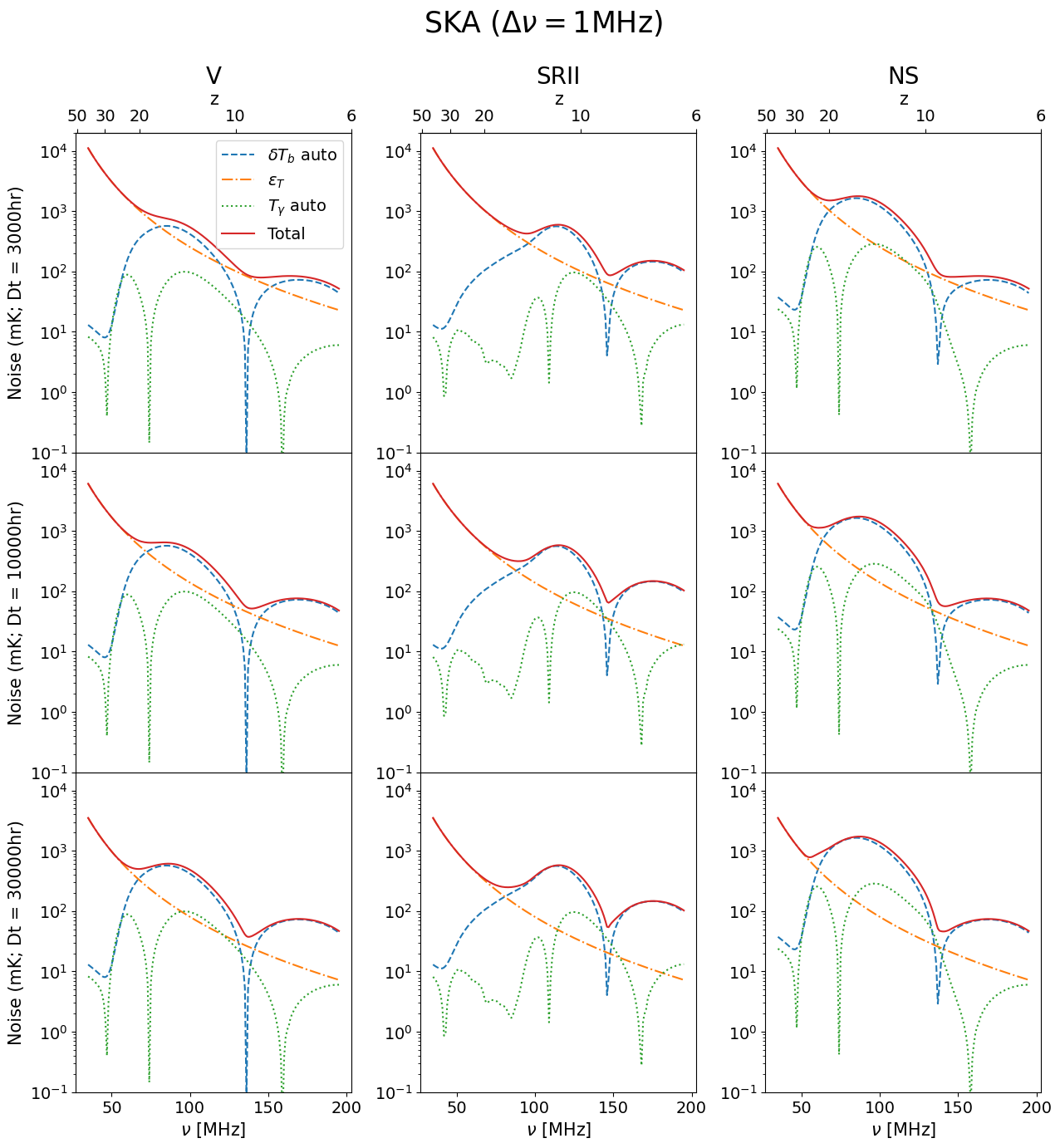}

\caption{Composition of the noise in the case of $\Delta\nu=1\,$MHz. Different
observing times are also investigated (top: 3000 hr; middle: 10000
hr; bottom: 30000 hr). The sky brightness ($\epsilon_{T}$: orange, dot-dashed)
dominates the net noise (red, solid) at $z\gtrsim15-20$, while the noise in the lower-redshift
epoch is dominated by the uncontrollable, intrinsic quantity $C_{\ell}^{\delta T_{b}\delta T_{b}}$
(blue, dashed).\label{fig:noise}}
\end{figure}

The observable $s_{\ell}(\bar{\nu})$ defined in Eq. (\ref{eq:Clratio}) of
the main text has the noise spectrum $\sigma_{s_{\ell}}(\bar{\nu})$
to be shown in Eq. (\ref{eq:sigma2}) below. We show how we obtained
the expression. First, one should start from the error propagation
(we drop the frequency dependence for readability): 
\begin{align}
  \sigma_{s_{\ell}}^{2}=&T_{\gamma,0}^{2}\left(\frac{C_{\ell}^{gT}}{C_{\ell}^{gT_{\gamma}}}\right)^{2} \nonumber \\
                          &\left(\frac{\sigma_{C_{\ell}^{gT}}^{2}}{\left(C_{\ell}^{gT}\right)^{2}}+\frac{\sigma_{C_{\ell}^{gT_{\gamma}}}^{2}}{\left(C_{\ell}^{gT_{\gamma}}\right)^{2}}-2\frac{\sigma_{C_{\ell}^{gT}C_{\ell}^{gT_{\gamma}}}}{C_{\ell}^{gT}C_{\ell}^{gT_{\gamma}}}\right),\label{eq:errorprop}
\end{align}
where we ignored the variance in the global CMB temperature $\sigma_{T_{\gamma,0}}^{2}=0.57^{2}\,{\rm mK}^{2}$,
which is much smaller than the net noise we find (see Fig. \ref{fig:noise}). Realizing
mutually related gaussian fields at given harmonics ($\ell$, $m$)
by combinations of independent gaussian fields are useful \citep{Cabre2007,Giannantonio2008}:
\begin{widetext}
\begin{align}
a_{\ell m}^{g}= & \sqrt{C_{\ell}^{gg}}\psi_{1,\ell m},\nonumber \\
a_{\ell m}^{T_{\gamma}}= & \frac{C_{\ell}^{gT_{\gamma}}}{\sqrt{C_{\ell}^{gg}}}\psi_{1,\ell m}+\sqrt{C_{\ell}^{T_{\gamma}T_{\gamma}}-\frac{(C_{\ell}^{gT_{\gamma}})^{2}}{C_{\ell}^{gg}}}\psi_{2,\ell m},\nonumber \\
a_{\ell m}^{T}= & \frac{C_{\ell}^{gT}}{\sqrt{C_{\ell}^{gg}}}\psi_{1,\ell m}+\left(C_{\ell}^{T_{\gamma}T}-\frac{C_{\ell}^{gT_{\gamma}}C_{\ell}^{gT}}{C_{\ell}^{gg}}\right)/\sqrt{C_{\ell}^{T_{\gamma}T_{\gamma}}-\frac{(C_{\ell}^{gT_{\gamma}})^{2}}{C_{\ell}^{gg}}}\psi_{2,\ell m}\nonumber \\
 & +\sqrt{C_{\ell}^{TT}-\frac{C_{\ell}^{T_{\gamma}T_{\gamma}}\left(C_{\ell}^{gT}\right)^{2}+C_{\ell}^{gg}\left(C_{\ell}^{T_{\gamma}T}\right)^{2}-2C_{\ell}^{gT_{\gamma}}C_{\ell}^{gT}C_{\ell}^{T_{\gamma}T}}{C_{\ell}^{gg}C_{\ell}^{T_{\gamma}T_{\gamma}}-\left(C_{\ell}^{gT_{\gamma}}\right)^{2}}}\psi_{3,\ell m},\label{eq:indepgauss}
\end{align}
\end{widetext}
where $\psi_{1,\ell m}$, $\psi_{2,\ell m}$, and $\psi_{3,\ell m}$
are mutually independent, unit-variance gaussian random fields for
harmonics ($\ell$, $m$) that satifsy
\begin{align}
\left\langle \psi_{i,\ell m}\psi_{j,\ell m}\right\rangle  & =\delta_{ij},\nonumber \\
\left\langle \psi_{i,\ell m}^{3}\psi_{j,\ell m}\right\rangle  & =3\delta_{ij},\nonumber \\
\left\langle \psi_{i,\ell m}^{2}\psi_{j,\ell m}^{2}\right\rangle  & =3\delta_{ij}+(1-\delta_{ij}),\label{eq:psi}
\end{align}
where $\left\langle \;\right\rangle $ denotes the ensemble average.
We also use the fact
\begin{align}
C_{\ell}^{gT} & =C_{\ell}^{gT_{\gamma}}+C_{\ell}^{g\delta T_{b}}=\left(1+\frac{\delta T_{21}}{T_{\gamma,0}}\right)C_{\ell}^{gT_{\gamma}},\nonumber \\
C_{\ell}^{TT} & =C_{\ell}^{T_{\gamma}T_{\gamma}}+C_{\ell}^{\delta T_{b}\delta T_{b}},\nonumber \\
C_{\ell}^{T_{\gamma}T} & =C_{\ell}^{T_{\gamma}T_{\gamma}},\label{eq:Cadd}
\end{align}
where we assumed $C_{\ell}^{T_{\gamma}\delta T_{b}}=0$ in the last
two equations, which is just an approximation because the evolving
Dopper effect during reionization can give a substantial cross-correlation
between the 21-cm background and CMB, especially so when reionization
proceeds fast, yet still subdominant to $C_{\ell}^{T_{\gamma}T_{\gamma}}$
\citep{Alvarez2006}. Also another cross-correlation comes through the
common ISW effect:
\begin{equation}
C_{\ell}^{T_{\gamma}\delta T_{b}}=\frac{C_{\ell}^{gT_{\gamma}}C_{\ell}^{g\delta T_{b}}}{C_{\ell}^{gg}}
\end{equation}
which is again quite subdominant to $C_{\ell}^{T_{\gamma}T_{\gamma}}$.

Using the equality
\begin{align}
\sigma_{C_{\ell}^{XY}}^{2} & =\left\langle a_{\ell m}^{X}a_{\ell m}^{Y*}a_{\ell m}^{X*}a_{\ell m}^{Y}\right\rangle -\left(C_{\ell}^{XY}\right)^{2},\nonumber \\
\sigma_{C_{\ell}^{XY}C_{\ell}^{XZ}} & =\left\langle a_{\ell m}^{X}a_{\ell m}^{Y*}a_{\ell m}^{X*}a_{\ell m}^{Z}\right\rangle -C_{\ell}^{XY}C_{\ell}^{XZ},\label{eq:sigequal}
\end{align}
together with Eqs. (\ref{eq:indepgauss} -- \ref{eq:sigequal}),
we obtain
\begin{widetext}
\begin{align}
\sigma_{C_{\ell}^{gT_{\gamma}}}^{2} & =\left(C_{\ell}^{gT_{\gamma}}\right)^{2}+C_{\ell}^{gg}C_{\ell}^{T_{\gamma}T_{\gamma}},\nonumber \\
\sigma_{C_{\ell}^{gT}}^{2} & =\left(C_{\ell}^{gT}\right)^{2}+C_{\ell}^{gg}C_{\ell}^{TT}=\left(1+\frac{\delta T_{21}}{T_{\gamma,0}}\right)^{2}\left(C_{\ell}^{gT_{\gamma}}\right)^{2}+C_{\ell}^{gg}\left(C_{\ell}^{T_{\gamma}T_{\gamma}}+C_{\ell}^{\delta T_{b}\delta T_{b}}\right)\nonumber \\
\sigma_{C_{\ell}^{gT_{\gamma}}C_{\ell}^{gT}} & =C_{\ell}^{gT_{\gamma}}C_{\ell}^{gT}+C_{\ell}^{gg}C_{\ell}^{T_{\gamma}T}=\left(1+\frac{\delta T_{21}}{T_{\gamma,0}}\right)\left(C_{\ell}^{gT_{\gamma}}\right)^{2}+C_{\ell}^{gg}C_{\ell}^{T_{\gamma}T_{\gamma}}.
\label{eq:sigs}
\end{align}
Finally, using Eqs. (\ref{eq:errorprop}) and (\ref{eq:sigs})
we obtain
\begin{equation}
\sigma_{s_{\ell}}^{2}(\bar{\nu})=\left\langle T_{\gamma,0}\right\rangle ^{2}\frac{\left(C_{\ell}^{gg}+\bar{n}_{g}^{-1}\right)\left\{ C_{\ell}^{\delta T_{b}\delta T_{b}}(\bar{\nu})+\epsilon_{\ell,T}(\bar{\nu})+\left(\frac{\delta T_{21}(\bar{\nu})}{\left\langle T_{\gamma,0}\right\rangle }\right)^{2}C_{\ell}^{T_{\gamma}T_{\gamma}}\right\} }{\left(C_{\ell}^{gT_{\gamma}}\right)^{2}f_{{\rm sky}}(2\ell+1)},\label{eq:sigma2}
\end{equation}
\end{widetext}
where systematic errors in auto-correlations are now included and
the number of modes with imperfect sky coverage is accounted for as
well. The net noise is given by
\begin{equation}
N^{-2}(\bar{\nu})=\sum_{\ell=\ell_{{\rm min}}}^{\ell_{{\rm max}}}\sigma_{s_{\ell}}^{-2}(\bar{\nu}).\label{eq:netnoise}
\end{equation}

Fig. 1 shows how the observing time affects the SNR.
In the case of $\Delta\nu=5\,$MHz, the gain from having $\Delta t=30000\,$hr
instead of $\Delta t=10000\,$hr is not much, because the sky brightness
noise is still too high at $z\sim30$ and the lower-redshift ($z\lesssim15-20$)
noise is intrinsic, coming mostly from the 21-cm auto-correlation
$C_{\ell}^{\delta T_{b}\delta T_{b}}(\bar{\nu})$. See the relative
noise contribution in Fig. \ref{fig:noise}. 
The even easier $\Delta t=3000\,$hr case also gives moderate sensitivity
during the early EoR phase, and thus should be considered seriously in
the early run of the SKA1-LOW.

\subsection{superstructure 21-cm ISW}
\label{sec:ssnoise}

If one were to use superstructures, this is not to use the power spectrum
as the full-sky cross-correlation scheme does. Instead, this is close
to the full imaging. Therefore, the automatic removal of the Milky
Way foreground does not apply to this strategy. Let us imagine that
we have the temperature field where all the foreground components
(CMB, Milky Way, extragalactic) are removed.

The noise is estimated as follows. Superstructures are characterized
by their large angular span, $\sim4^{\circ}$ in radius, and the temperature
field is convoluted with specific smoothing or matching filters $\psi^{{\rm MF}}(\left|\hat{n}-\hat{n}_{0}\right|)=\psi^{{\rm MF}}(\theta)$
where $\hat{n}_{0}$ is the location of the center of a superstructure
and $\theta$ is the angle away from $\hat{n}_{0}$. For example,
\citep{Granett2008} used a smoothing filter that averages the temperature
in the region where $\theta<\theta_{{\rm th}}$ and further subtracts
the average temperature where $\theta_{{\rm th}}<\theta<\sqrt{2}\theta_{{\rm th}}$,
and claimed the oddly strong ($A_{{\rm ISW}}\simeq5$) ISW effect
on superstructures. We follow \citep{Nadathur2016} to estimate the
noise: we take three different matching filters shown in \citep{Nadathur2016}
that would each give the filtered 21-cm temperature field
\begin{equation}
\delta T_{b}'(\hat{n}',\bar{\nu})=\int d\hat{n}\,\delta T_{b}(\hat{n},\bar{\nu})\psi^{{\rm MF}}(\left|\hat{n}'-\hat{n}\right|).\label{eq:filtered}
\end{equation}
The variance of this filetered field will be given by
\begin{equation}
\sigma_{\delta T_{b}}^{2}(\bar{\nu})=\sum_{\ell}\left(C_{\ell}^{\delta T_{b}\delta T_{b}}+\epsilon_{\ell,T}\right)\left(\psi_{\ell0}^{{\rm MF}}\right)^{2},\label{eq:sigfiltered}
\end{equation}
which is synonymous to the mass variance of filtered density field,
$\sigma_{M}$. Here, $\psi_{\ell0}^{{\rm MF}}$ is the expansion coefficient
in Legendre polynomials of $\psi^{{\rm MF}}(\theta)$:
\begin{equation}
\psi^{{\rm MF}}(\theta)=\sum_{\ell}\psi_{\ell0}^{{\rm MF}}Y_{\ell0}(\cos\theta).\label{eq:legendre}
\end{equation}
We found that even with the difference in the three filters, the variance
calculated by Eq. (\ref{eq:sigfiltered}) all lead to about the
same values at all frequencies. We inverted Eq. (\ref{eq:legendre})
using orthogonality of $Y_{\ell0}$, and calculated the integral to
obtain $\psi_{\ell0}^{{\rm MF}}$ in a brute-force way. We took adaptively
fine samples in the Legendre polynomials, which was somewhat computationally
costly. Because $\psi^{{\rm MF}}$ is a low-$\ell$-pass filter, we
obtained $\psi_{\ell0}^{{\rm MF}}$ only up to $\ell=200$ and set
the rest to zero. We found that $\psi_{\ell0}^{{\rm MF}}$ obtained
this way reproduces $\psi^{{\rm MF}}(\theta)$ well enough.

Because multiple superstructure ISW signals are stacked, if there
are $N_{{\rm ss}}$ superstructures stacked then the final noise is
given by
\begin{equation}
\sigma_{\delta T_{b}}^{2}(\bar{\nu})=\sum_{\ell}\left(C_{\ell}^{\delta T_{b}\delta T_{b}}+\epsilon_{\ell,T}\right)\left(\psi_{\ell0}^{{\rm MF}}\right)^{2}/N_{{\rm ss}},\label{eq:reducedsig}
\end{equation}
and the signal is given by Eq. (\ref{eq:super21}) of the main text. We used
Eq. (\ref{eq:reducedsig}) to calculate the noise with $N_{{\rm ss}}=200$,
twice as large as what was used in \citep{Granett2008} or about 1/2
of that in \citep{Nadathur2012}, and obtained Fig. \ref{fig:ssSNR} of the main
text.


\end{document}